\documentclass{webofc}

\usepackage[varg]{txfonts}   
\usepackage{hyperref}
\usepackage{url}
\hypersetup{colorlinks=true,citecolor=blue,urlcolor=blue,linkcolor=blue}
\begin{document}
\title{What happens in hydrodynamic simulations of heavy-ion collisions when causality is violated?}
%
%

\author{\firstname{Lorenzo} \lastname{Gavassino}\inst{1}\fnsep\thanks{\email{lorenzo.gavassino@gmail.com}} \and
        \firstname{Henry} \lastname{Hirvonen}\inst{1}\fnsep\thanks{\email{henry.v.hirvonen@vanderbilt.edu}} \and
        \firstname{Jean-Fran\c{c}ois} \lastname{Paquet}\inst{2,1}\fnsep\thanks{\email{jean-francois.paquet@vanderbilt.edu }}
\and
        \firstname{Mayank} \lastname{Singh}\inst{2}\fnsep\thanks{\email{mayank.singh@vanderbilt.edu }}
\and
        \firstname{Gabriel} \lastname{Soares Rocha}\inst{3,2}\fnsep\thanks{\email{gabrielsr@id.uff.br}}
}

\institute{Department of Mathematics, Vanderbilt University, Nashville TN, USA
\and
Department of Physics and Astronomy, Vanderbilt University, Nashville TN, USA 
\and
Instituto de F\'isica, Universidade Federal Fluminense, Niter\'oi, Brazil
          }

\abstract{We summarize our recent investigations on how causality violations in Israel-Stewart-type relativistic viscous hydrodynamic simulations can give rise to both analytical and numerical instabilities. The classification of spacetime regions into \emph{causal and stable} (``good''), \emph{acausal but stable} (``bad''), and \emph{acausal and unstable} (``ugly'') is reviewed. We compare the predictions of the MUSIC hydrodynamic solver with an analytical solution, and demonstrate how the acausality-driven instabilities develop in a simple one-dimensional scenario.
}
\maketitle
\section{Introduction}

Relativistic viscous hydrodynamics plays a key role in modeling the quark-gluon plasma created in ultra-relativistic heavy-ion collisions \cite{FlorkowskiReview2018}. One of the most widely used formulations is Israel-Stewart (IS) theory \cite{Israel_Stewart_1979}, which improves upon relativistic Navier-Stokes by introducing relaxation times (say, \( \tau_\Pi \)) for dissipative quantities such as the bulk pressure \( \Pi \), giving
\begin{equation}
\tau_\Pi u^\mu \partial_\mu \Pi +\Pi =-\zeta \partial_\mu u^\mu \, \quad \quad (\zeta=\text{``bulk viscosity''},\,\,\, u^\mu =\text{``flow velocity''}).
\end{equation}
While IS theory is designed to be causal and stable near equilibrium \cite{Hishcock1983}, recent studies have revealed that this is not always the case in far-from-equilibrium regimes \cite{Causality_bulk,BemficaCausalityShear2020xym}. Specifically, it has been shown that the speed of propagation of information in the fluid description can exceed the speed of light at early times in state-of-the-art simulations \cite{Plumberg2022}, meaning that the principle of causality may be violated.

Since causality is deeply related to stability and well-posedness \cite{GavassinoCausality2021,GavassinoSuperluminal2021}, its violations raise fundamental doubts concerning the applicability (and meaningfulness) of the Israel-Stewart equations at high gradients. To gain better insight into this problem, we set out to explore the connection between superluminal signal propagation and instabilities, both physical and numerical. This is done by comparing the behavior of an analytical solution of Israel-Stewart theory that exhibits a transition between stable and unstable regimes with the corresponding numerical solution calculated using the MUSIC code. A more extensive analysis can be found in \cite{GavassinoSimulations2025bsn}.

\section{Classification of fluid cells}

\begin{figure}
    \centering
\includegraphics[width=0.27\linewidth]{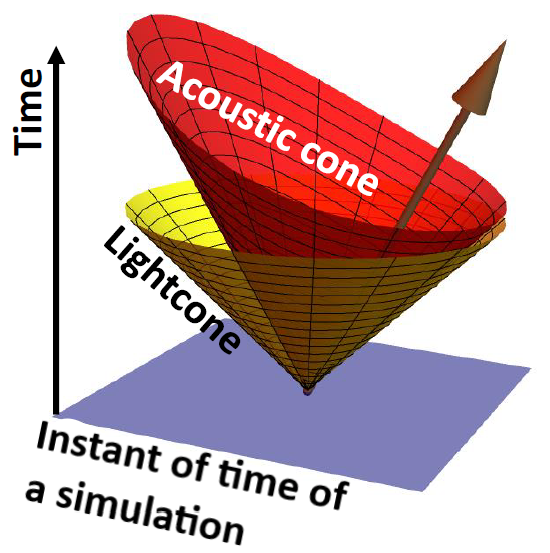}
\includegraphics[width=0.32\linewidth]{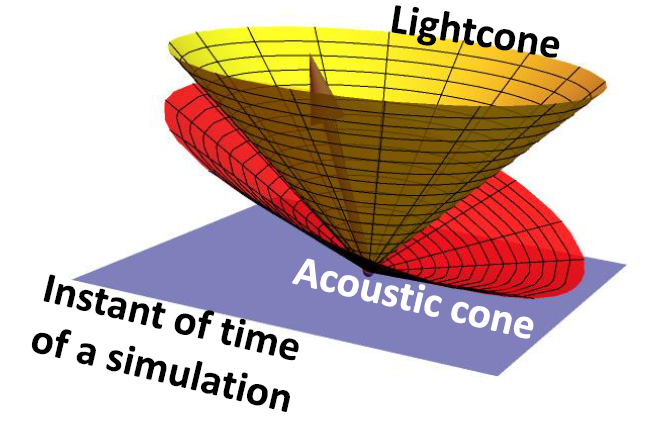}
\includegraphics[width=0.30\linewidth]{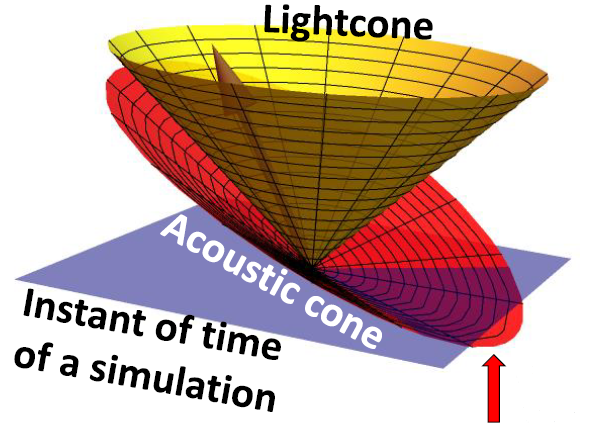}
\caption{Spacetime diagram illustrating the acoustic geometry of a fluid element in a ``good'' (left panel), ``bad'' (middle panel), and ``ugly'' (right panel) state. In the good state, the acoustic cone is contained within the lightcone ($0\leq w\leq 1$). In a bad state, the acoustic cone exits the lightcone ($w>1$), but still points to the future relative to the observer who is solving the equation ($vw<1$). In the ugly state, a portion of the acoustic cone points to the past ($vw\geq 1$), or there is no acoustic cone  ($w^2<0$).}
    \label{fig:goodbadugly}
\vspace{-0.5cm}    
\end{figure}

 In Israel-Stewart theory with only bulk viscosity (for clarity), perturbations propagate along characteristics defined by an ``acoustic cone'' whose geometry is encoded by an effective acoustic metric \cite{Causality_bulk}: $G_{\mu \nu}=-u_\mu u_\nu +\dfrac{1}{w^2} (g_{\mu \nu}+u_\mu u_\nu)$.
The quantity \( w \) is the speed at which information propagates in the local rest frame of the substance, which for bulk viscous fluids at zero chemical potential is given by
\begin{equation}\label{wcs2}
w^2= c_s^2 +\dfrac{\zeta}{\tau_\Pi (\varepsilon+P+\Pi)} \, ,
\end{equation}
with $c_s^2$ the sound speed, $\varepsilon$ the energy density, and $P$ the pressure. This acoustic geometry allows us to identify three distinct regimes in which a fluid cell can be found, depending on the value of $w$ and on the speed of fluid motion $v$ in the simulation frame (see figure \ref{fig:goodbadugly}):
\begin{itemize}
    \item \textbf{Good:} Causal, and therefore stable (\( w^2 \in [0,1] \)).
    \item \textbf{Bad:} Acausal but stable in the reference frame of the simulation (\( w^2 > 1 \), but \( v w < 1 \)).
    \item \textbf{Ugly:} Acausal and unstable (\( w^2 < 0 \) or \( v w \geq 1 \)).
\end{itemize}
The distinction between ``bad'' and ``ugly'' arises from the fact that causality violations do not automatically lead to instabilities in all reference frames. Instead, instabilities arise in a given Lorentz frame whenever part of the acoustic cone is past-directed,  i.e. when signals can propagate from the future to the past. This occurs when \( v w \geq 1 \). In this case, dissipation can work in reverse, amplifying perturbations instead of damping them.

More severe pathologies occur when \( w^2 < 0 \), in which case the system becomes elliptic. Elliptic equations do not possess a well-posed initial value formulation and admit modes that grow arbitrarily fast with wavenumber. These are known as Hadamard instabilities \cite{Rauch_book}. 


\begin{figure}[h!]
    \centering
\includegraphics[width=0.75\linewidth]{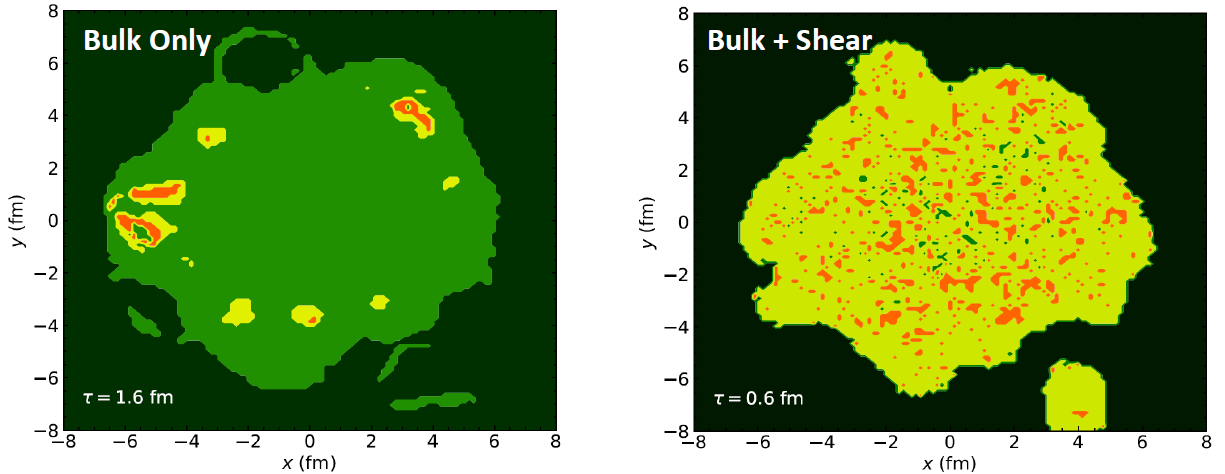}
\caption{Separation of fluid cells into good (green), bad (yellow), and ugly (red) in a simulation with only bulk viscosity (left panel) and both bulk and shear viscosities (right panel). 
The energy-momentum tensor is initialized using the IP-Glasma model in both cases.}
    \label{fig:2D}
\vspace{-0.5cm}        
\end{figure}

Figure \ref{fig:2D} shows how good, bad, and ugly regions tend to distribute in realistic heavy-ion simulations with IP-Glasma initial data.\footnote{The input file employed is publicly available at \url{https://github.com/MUSIC-fluid/MUSIC/blob/public_stable/example_inputfiles/IPGlasma_2D/input/epsilon-u-Hydro-t0.6-0.dat}.} Due to the large gradients, the early stages of a simulation tend to be populated by both bad and ugly cells. 
As soon as ugly cells develop, we expect the simulation to develop instabilities and no longer yield an accurate solution of the differential equation. In practice, regulators used in Israel-Stewart solvers and UV cutoffs from the numerical grid may play a role in limiting these instabilities.

\section{Numerical tests with a benchmark Solution}

To better understand the onset of instabilities in the mathematical solution as opposed to the numerical solution, we consider a homogeneous fluid with only bulk viscosity and equation of state \( \varepsilon=3P \), undergoing the following flow (with $P_0$ and $t_0$ arbitrary constants):
\begin{equation}\label{solviamo}
\begin{split}
u^\mu={}&\dfrac{1}{\sqrt{1-v^2}}(1,v,0,0)\, , \quad
P={} \dfrac{1}{3} (2{-} v)P_0 \, ,\quad
\Pi={} \bigg[ \dfrac{1}{v}+ \dfrac{v}{3} -\dfrac{8}{3} \bigg] P_0  \, , \\
t={}& t_0+ \frac{1}{312} \Bigg[ 78 u^0 -\sqrt{39 \left(1586 \sqrt{13}-5155\right)} \arctan\left(\frac{\sqrt{\frac{1}{2} \left(\sqrt{13}-1\right)} }{(1+v)u^0}\right) \\
& -624 \, \text{arctanh} \left(\frac{1}{u^0(1+v)}\right)+\sqrt{39 \left(1586 \sqrt{13}+5155\right)} \text{arctanh}\left(\frac{\sqrt{\frac{1}{2} \left(\sqrt{13}+1\right)} }{(1+v)u^0}\right) \Bigg] \, . \\
\end{split}
\end{equation}
This is an analytical solution of the Israel-Stewart equations with bulk viscosity alone, with $\tau_\Pi=1$ (in code units) and $\zeta=P\tau_\Pi$. Such solution bifurcates into two branches at flow speed \( v\approx 0.838 \)  (where \( v w = 1 \)). One branch ($v<0.838$) evolves from bad to good and reaches equilibrium at late times. The other branch is entirely ugly, and it exhibits unstable behavior.

\begin{figure}
    \centering
\includegraphics[width=0.4\linewidth]{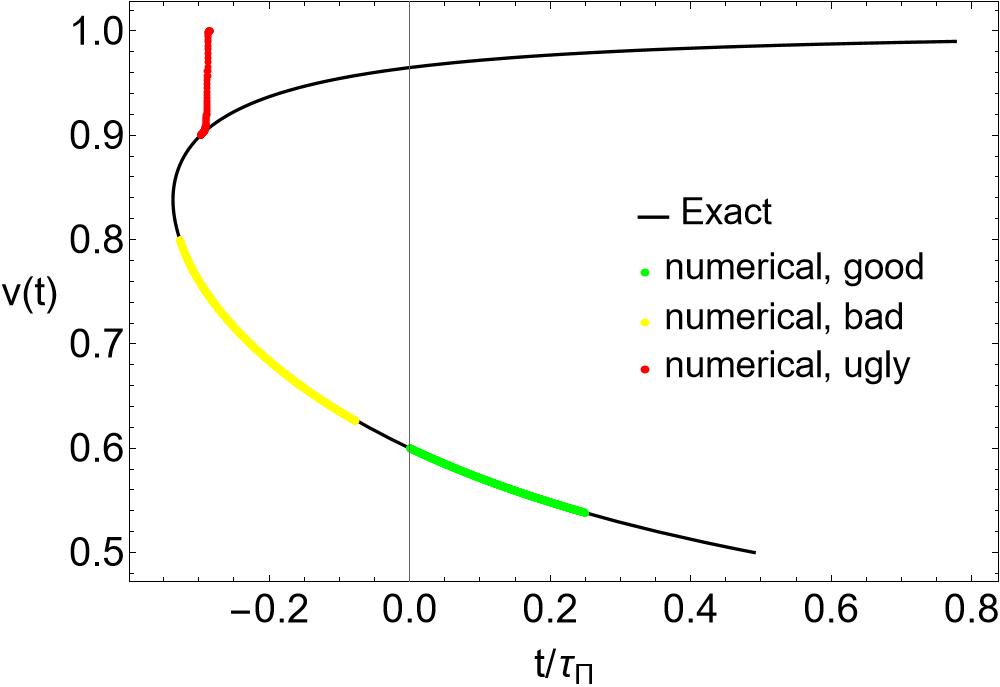}
\includegraphics[width=0.4\linewidth]{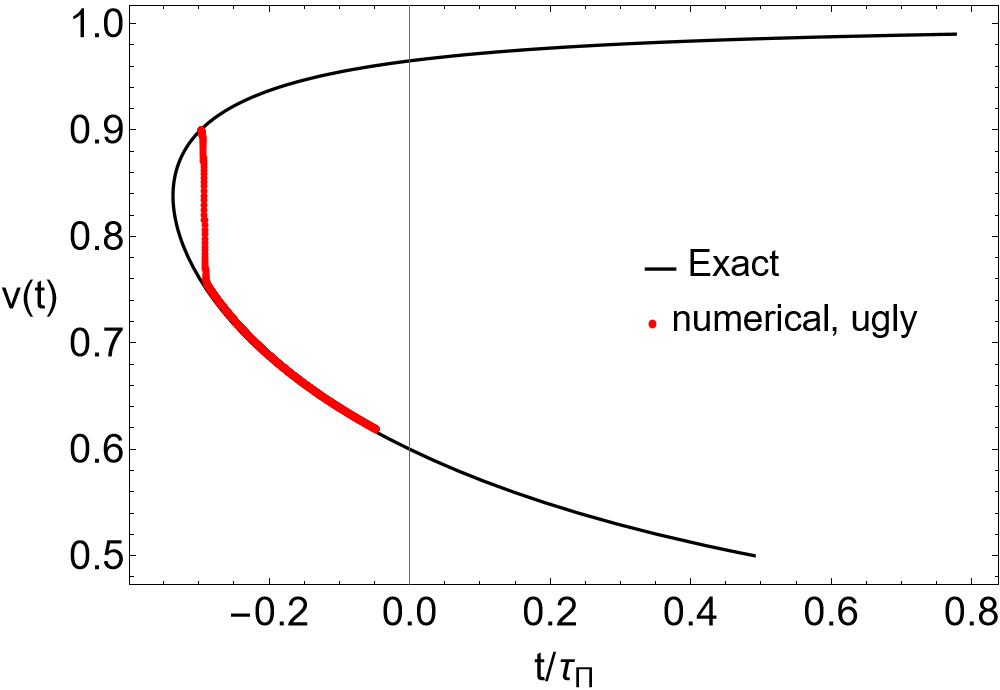}
\caption{(Left panel) Numerical solution of a homogeneous, Israel-Stewart fluid with only bulk viscosity, with good (green), bad (yellow), and ugly (red) initial conditions, respectively, contrasted with the exact solution \eqref{solviamo} (black). (Right panel) Ugly solution with zero-time-gradient  initial conditions \cite{GavassinoSimulations2025bsn}.}
    \label{fig:expl}
    \vspace{-0.5cm}    
\end{figure}

Using the MUSIC solver, we simulate the benchmark solution above with initial conditions spanning the good, bad, and ugly regimes (see figure \ref{fig:expl}). We find perfect agreement with the analytical solution in the good and bad regimes. In the ugly regime, the simulation exhibits runaway behavior, but only if we initialize the time derivatives of the source consistently with the equations of motion (left panel). Instead, the zero-time-gradient initialization (see Ref.~\cite{GavassinoSimulations2025bsn}, for a discussion), employed by default in some numerical procedures, suppresses the instability (right panel).

\section{Conclusions}

Causality violations in relativistic hydrodynamics are not merely theoretical curiosities; they have tangible effects on both physical evolution and numerical simulations. Using a benchmark analytical solution, we demonstrated how instabilities emerge precisely at the threshold \( v w = 1 \). In this highly simplified scenario, numerical simulations confirm this behavior, and show that careful modeling and initialization can prevent unphysical outcomes. These findings provide critical guidance for future implementations of relativistic hydrodynamics in high-energy physics.

\section*{Acknowledgments}

H.H., J.-F.~P., M.~S. and G.~S.~R. are supported by Vanderbilt University and by the U.S. Department of Energy, Office of Science under Award Number DE-SC-0024347.
H.H. is also partly supported by the U.S. Department of Energy, Office of Science under Award Number DE-SC0024711, and the National Science Foundation under Grant No. DMS-2406870. L.G. is partially supported by a Vanderbilt Seeding Success grant. 

\bibliography{Biblio}

\label{lastpage}

\end{document}